\documentclass[12pt,a4paper]{article}

\usepackage{jheppub} 
\usepackage[english]{babel}
\usepackage[utf8]{inputenc}
\usepackage[T1]{fontenc} 

\usepackage{graphicx}
\usepackage{dcolumn}
\usepackage{bm}

\newcommand{\beq}{\begin{equation}}
\newcommand{\eeq}{\end{equation}}
\newcommand{\bea}{\begin{eqnarray}}
\newcommand{\eea}{\end{eqnarray}}

\newcommand{\bj}{\bm{j}}
\newcommand{\bk}{\bm{k}}
\newcommand{\calA}{\mathcal{A}}
\newcommand{\calF}{\mathcal{F}}
\newcommand{\calJ}{\mathcal{J}}
\newcommand{\calW}{\mathcal{W}}
\newcommand{\bcalB}{\bm{\mathcal{B}}}
\newcommand{\bcalE}{\bm{\mathcal{E}}}
\newcommand{\bcalD}{\bm{\mathcal{D}}}
\newcommand{\bcalH}{\bm{\mathcal{H}}}
\newcommand{\bcalJ}{\bm{\mathcal{J}}}
\newcommand{\bcalM}{\bm{\mathcal{M}}}

\def\d{\mathrm{d}}

\def\beq{\begin{equation}}
\def\eeq{\end{equation}}
\def\bdm{\begin{displaymath}}
\def\edm{\end{displaymath}}

\def\bea{\begin{eqnarray}}
\def\eea{\end{eqnarray}}

\makeatletter
\newcommand{\pushright}[1]{\ifmeasuring@#1\else\omit\hfill$\displaystyle#1$\fi\ignorespaces}
\newcommand{\pushleft}[1]{\ifmeasuring@#1\else\omit$\displaystyle#1$\hfill\fi\ignorespaces}

\makeatother

\graphicspath{{figures/}{graphics/}{../../../../graphics/}{../../../graphics/}{../../graphics/}{../graphics/}}

\begin{document}

\title{Holographic Plasmons}


\author[a]{U.~Gran,}
\author[a]{M.~Torns\"o}
\author[b]{and T.~Zingg}


\affiliation[a]{Department of Physics,
Division for Theoretical Physics,
Chalmers University of Technology\\
SE-412 96 G\"{o}teborg,
Sweden}
\affiliation[b]{Nordita,
Stockholm University and KTH Royal Institute of Technology\\
Roslagstullsbacken 23,
SE-106 91 Stockholm,
Sweden}

\emailAdd{ulf.gran@chalmers.se}
\emailAdd{marcus.tornso@chalmers.se}
\emailAdd{zingg@nordita.org}

\abstract{
Since holography yields exact results, even in situations where perturbation theory is not applicable, it is an ideal framework for modeling strongly correlated systems. 
We extend previous holographic methods to take the dynamical charge response into account and use this to perform the first holographic computation of the dispersion relation for plasmons. 
As the dynamical charge response of strange metals can be measured using the new technique of momentum-resolved electron energy-loss spectroscopy (M-EELS), plasmon properties are the next milestone in verifying predictions from holographic models of new states of matter.


}

\maketitle


\section{Introduction}
\label{sec:intro}


A ubiquitous phenomenon in physical systems with free charge carriers is \emph{plasma oscillations}. These collective modes arise when the charge density fluctuates and induces a polarization which in turn backreacts on the charge density, trying to restore it to its original configuration. If perturbed by the resonant frequency, a self-sustained propagating wave of charge transport is created. The quanta of these collective excitations, sometimes also the excitations themselves, are called \emph{plasmons}.


These plasmons have many technical applications. 
Already in Roman times cups were colored via plasmonic effects, though a more well-known application is probably the staining of church windows during the Middle Ages. 
Modern applications in the field of \emph{plasmonics}~\cite{Maier:2007:10.1007/0-387-37825-1} include extremely sensitive devices for biosensing and using plasmons for communication within circuits. One key feature in plasmonics is that plasmons have a shorter wavelength compared to an electromagnetic wave of the same frequency, up to a factor of $\alpha^{-1}\sim 100$ for graphene~\cite{2011NanoL..11.3370K}, where $\alpha$ is the fine structure constant. This phenomenon, called \emph{wave localization}, could, among other things, allow for the miniaturization of circuits based on plasmonics compared to other technologies~\cite{SPSO2003,ZIA200620}.


The relevant quantity to consider when studying plasmons is the dielectric function $\varepsilon(\omega,k)$ as it takes into account the dynamical polarization of the medium. Plasmons are then obtained as longitudinal solutions to the equation $\varepsilon(\omega,k)=0$, which we will refer to as the \emph{plasmon condition}.
When studying plasmons using standard QFT methods, one has to resort to various approximations, e.g.~the random phase approximation (RPA), to extract the relevant parts of the higher loop contributions to the dielectric function. One important feature of holographic duality is that it yields the exact expressions for the dielectric function, i.e.~including loop contributions to all orders. This makes holographic duality a powerful framework to use in order to analyze the dynamical response of a system, and in particular for studying plasmons.



An additional motivation to study plasmons via holographic duality is recent advances in experimental techniques targeting `strange metals'~\cite{2017arXiv170801929M}, a phase existing above the critical temperature, $T_c$, in many high-temperature superconductors.
As superconductivity is suppressed in this phase, having a better understanding of it is crucial to develop methods to increase $T_c$.
Furthermore, it has been linked to strongly coupled quantum critical behaviour~\cite{vandeMarel:2003wn,Cooper603} and displays the absence of long-lived quasi-particle excitations in spectral probe studies. These features makes the strange metal phase a prime target to study using holographic duality. With the new method of momentum-resolved electron energy-loss spectroscopy (M-EELS) the dynamical charge response of the strange metal $\rm{Bi_{2,1}Sr_{1,9}CaCu_{2}O_{8+x}}$ (BSCCO) has been successfully measured, in particular the plasmon energy and lineshape for small momenta \cite{2017arXiv170801929M}. This opens up a novel way to compare holographic results on strange metals, in particular those concerning plasmons, to experiments. We will comment on the agreement between the experimental results and a holographic model in section~\ref{sec:RNmetal}.


During the last decades holographic duality has been established as a valuable tool that provides a new perspective on strongly coupled phenomena in condensed matter physics -- for an introduction to the subject see e.g.~\cite{Ammon:2015wua,Zaanen:2015oix,Hartnoll:2016apf}.
In a quantum field theory (QFT) describing a condensed matter system, the renormalization group scale can be geometrized, adding an extra ``energy'' dimension to the theory. 
The highly non-trivial claim of holographic duality is that the dynamics of this extended spacetime is governed by a \emph{gravity} theory. 
The dual QFT can be viewed as living on the \emph{boundary} of the extended spacetime, which is called the \emph{bulk} spacetime. 
What makes holographic duality so useful is that it maps strongly coupled physics in the QFT to weakly coupled physics in the gravity theory, and vice versa, thus providing a new method to analyze and understand strongly coupled phenomena, e.g.~in condensed matter physics.

The original motivation for the duality came from string theory~\cite{Maldacena:1997re,Gubser:1998bc,Witten:1998qj}, but it is now believed to hold much more generally and can e.g.~be derived purely within ordinary gravity theories~\cite{Fischetti:2012rd}. Furthermore, when additional symmetries are present, allowing computations on both sides of the duality to be performed and compared, the duality has so far passed all tests, see e.g.~\cite{Ammon:2015wua}.


The main result of this paper is to demonstrate how the holographic approach can be employed to establish a consistent description of plasmons for strongly correlated systems.
With the holographic dictionary, electromagnetic response in the boundary QFT is expressed in terms of solutions to a set of coupled PDEs, describing combined metric and gauge field fluctuations in the bulk.
The crucial ingredient is to identify the proper boundary conditions at spatial infinity that correspond to plasmons, i.e.~charge density fluctuations.
This is explained in section~\ref{sec:holoplasmons}. It turns out that the plasmon boundary conditions reduce to a particular choice of Robin boundary conditions, making it rather straightforward to apply the well-established techniques that have been developed in holographic duality.
This type of mixed boundary conditions can also be seen as a double trace deformation in the QFT~\cite{Witten:2001ua,Mueck:2002gm}, corresponding to an RPA form of the Green's function~\cite{Zaanen:2015oix}, meaning that our approach is consistent with conventional CMT~\footnote{We are grateful to Koenraad Schalm for bringing this point to our attention.}.
In addition, it also needs to be stressed the holographic computation of the conductivity, and hence the dielectric function, includes the full $\omega$ and $k$ dependence.
Thus, it corresponds to what is sometimes called the \emph{non-local} response, see e.g.~\cite{Wenger2017}, which takes into account the internal spreading of energy via the moving charge carriers. 



The reader familiar with holography may notice some similarity with the procedure of computing quasi-normal modes (QNM), i.e.~how perturbations of the bulk fields decay in time.
The crucial difference to keep in mind here is that while the system of PDEs in both cases originate from perturbing around a fixed background, the boundary conditions imposed at spatial infinity are fundamentally different.
The QNMs are identified with poles in holographic correlation functions and as such correspond to a situation where all sources have to vanish, which is synonymous with imposing Dirichlet conditions for all bulk fluctuations.
Plasmons are, however, not caracterized by poles in a holographic correlation function, since they depend in an essential way on the dynamical polarization of the material -- they are poles of a density-density response function to an \emph{external} field~\cite{nozieres1966theory}, and correspond to a situation where the response in the displacement current vanishes despite a non-zero electric field strength.
Thus, they always need a non-vanishing source to be present, which is why the corresponding boundary conditions are so different to those found in QNM calculations. Note that these sources are to be viewed as internal, being a consequence of the dynamical polarization.



\section{Boundary Electrodynamics}
\label{sec:holographic_em}

The holographic dictionary identifies a field strength $\calF$ and a conserved electric current $\calJ$ for the boundary QFT from bulk quantities,
\begin{eqnarray}
\calF	\;=\;	\frac{1}{\sqrt{\lambda}}\,F\bigr|_{\partial M}	\, ,	\quad
\calJ	\;=\;	\sqrt{\lambda}\,\imath_n W\bigr|_{\partial M}	\, ,
\label{eq:holodict}
\end{eqnarray}
where $F$ and $W$ are the bulk electromagnetic field strength and induction tensor and $n$ the normal direction on the boundary. We have also introduced $\lambda$, which we will identify with the parameter that relates the coupling strength for electromagnetism in the boundary theory with the one in the bulk as they do not have to be equal. The relative scaling in $\lambda$ for $\calF$ and $\calJ$ is determined due to them being related via conjugate quantities. The coupling strength for electromagnetism in the boundary theory has been absorbed in the fields, and can easily be reinstated by rescaling $\calJ \rightarrow e \calJ$ and $\rho \rightarrow e \rho$, where $e$ is the boundary coupling strength. Note, however, that the boundary physics is independent of $e$; it only depends on the relative strength of the bulk and the boundary coupling parameterized by $\lambda$ as we will elaborate on below.

The boundary induction tensor $\calW$ is then identified through
\begin{eqnarray}
\calJ	\;\;=\;\;	- \langle \rho \rangle \, dt + \bj
		\;\;=\;\; 	\star^{-1}d \star (\calF-\calW)
\, .
\label{eq:J_int}
\end{eqnarray}
This leads to the generic equations of motion for electromagnetism on the boundary,
\begin{eqnarray}
d \calF	\;=\;	0	\, , \quad
d \star \calW	\;=\;	\star\calJ_{ext}	\, ,
\label{eq:em_eom}
\end{eqnarray}
where $\calJ_{ext}$ is the current from external sources.
These equations are standard for a $U(1)$ gauge theory, independent of details of interaction or coupling with other fields.
Therefore, the standard decomposition of electromagnetic quantities -- see e.g.~\cite{nozieres1966theory} -- can be applied to the boundary field theory,
\begin{eqnarray}
\calF\! =\! \bcalE \!\wedge\! dt + \star^{-1} (\bcalB \!\wedge\! dt)	\, ,\;	
\calW\! =\! \bcalD \!\wedge\! dt + \star^{-1} (\bcalH \!\wedge\! dt)	. \qquad
\label{eq:fieldstrength_decomp}
\end{eqnarray}
From this, conductivity and dielectric functions, which both, in general, are tensorial quantities, are defined via the response to an electric field,
\begin{eqnarray}
\bj \;=\; \sigma \cdot \bcalE	\, , \qquad
\bcalD \;=\; \varepsilon \cdot \bcalE
\, .
\label{eq:cond+diel}
\end{eqnarray}
In a Minkowski background, the Maxwell equations~\eqref{eq:em_eom} can be cast into the 'classical' form,
\begin{align}
&\mathrm{div}\,\bcalB \;=\; 0	\, , \quad
&&\mathrm{div}\,\bcalD \;=\; \rho_{ext} \, , \quad    \nonumber \\
&\mathrm{curl}\,\bcalE \;=\; -\dot{\bcalB}	\, , \quad
&&\mathrm{curl}\,\bcalH \;=\; \bj_{ext} + \dot{\bcalD}	
\, . \quad
\label{eq:classical_maxwell}
\end{align}
Finding relations between various material constants is now a simple consequence of the respective definition and the equations of motion -- see e.g.~\cite{nozieres1966theory}.
After inserting~\eqref{eq:cond+diel} into~\eqref{eq:classical_maxwell}, Fourier-transforming and projecting onto $\bk$, it is a straightforward calculation to find,
\begin{eqnarray}
\bk \cdot \left( \varepsilon - 1 + \frac{\sigma}{i \omega} \right) \cdot \bcalE	\;\;=\;\; 	0
\, .
\end{eqnarray}
Thus, for the longitudinal mode, chosen in $x$ direction,
\begin{eqnarray}
\varepsilon_{xx} &=&	1 - \frac{\sigma_{xx}}{i \omega}
\, .
\label{eq:diel_00}
\end{eqnarray}
This allows to connect the dielectric function to the conductivity, which can easily be obtained from the Green's functions using the Kubo formula, i.e.
\begin{eqnarray}
\sigma_{xx} &=& \frac{i \omega }{k^2} \, \langle \rho \, \rho \rangle ~.
\label{eq:kubo}
\end{eqnarray}

In this paper, we will only consider backgrounds with no magnetic flux, but for completeness, we also mention that the other components of the dielectric function could be found from the more general relation,
\begin{eqnarray}
\left( \varepsilon - 1 + \frac{\sigma}{i \omega} \right) \cdot \bcalE	\;\;=\;\; 	\frac{\bk \times \bcalM}{\omega}
\, ,
\label{eq:diel_1}
\end{eqnarray}
with the magnetization $\bcalM	= \bcalB - \bcalH$. This is also valid for general dimensions when setting $\bk \times \bcalM := \star \,\bk \wedge \bcalM$.

\section{Dielectric Function and Plasmon Condition from Holography}
\label{sec:holoplasmons}

The longitudinal dielectric function $\varepsilon_{xx}$ has previously been studied holographically, for example in the context of holographic optics~\cite{Amariti:2010jw}, strongly coupled neutral plasmas~\cite{Forcella:2014dwa} and the wake potential in strongly coupled plasmas~\cite{Liu:2015sea}. These works generally employed~\eqref{eq:diel_00} to relate $\varepsilon_{xx}$ to the longitudinal conductivity~$\sigma_{xx}$. 

As shown in the previous section, this relation is a mere consequence of the definition of the holographic induction tensor via~\eqref{eq:J_int}. This makes it straightforward to identify plasmons by focusing on the longitudinal direction -- transverse collective modes do not give rise to charge density fluctuations and have therefore no relevance here.
By definition, charge density fluctuations are characterized by having vanishing displacement field~$\bcalD_x$ but non-zero electric field~$\bcalE_x$ when external sources are absent, i.e.~$\calJ_{ext}=0$. 
From equations~\eqref{eq:J_int} and~\eqref{eq:fieldstrength_decomp} follows that for plasmons,
\begin{eqnarray}
\dot{\bcalE}_x + \bj_x	&=&	0
\, .
\label{eq:plasmoncond_1}
\end{eqnarray}
Interestingly, this condition can be entirely expressed in terms of the quantities defined directly via the holographic correspondence~\eqref{eq:holodict}, without the need to explicitly compute the displacement field~$\bcalD$. In a homogeneous background, using the gauge choice $\calA_t=0$, \eqref{eq:plasmoncond_1} can also be written as
\begin{eqnarray}
\omega^2 \calA_x + \bcalJ_x	&=&	0
~.
\label{eq:plasmoncond_2}
\end{eqnarray}
In such a background, also the dielectric function~$\varepsilon$ can be defined, and~\eqref{eq:plasmoncond_2} can be shown to be equivalent to
\begin{eqnarray}
\varepsilon(\omega,k)	&=&	0~.
\label{eq:epscondition}
\end{eqnarray}
It is worth mentioning that this condition has more modes as solutions than what is commonly referred to as the plasmon mode, i.e.~a self-sourcing propagating oscillation.
If this condition was not fulfilled, then $\bcalD_x=0$ would imply
\begin{eqnarray}
\bcalE_x	\;=\;	\langle \rho \rangle
			\;=\;	0~,
\label{eq:QNMconditions}
\end{eqnarray}
meaning there would be no electric field present.

To calculate the holographic plasmon mode it is imperative to translate the condition~\eqref{eq:plasmoncond_1}, respectively~\eqref{eq:epscondition}, into a boundary condition for bulk fluctuations.
By construction, plasmons are found by studying response to a change in the electric field.
Thus, we fix boundary conditions such that only the perturbation $\delta\!A_x$ has a non-zero value at spatial infinity, to avoid overlap with response from fluctuations from other modes or channels. In particular, having non-zero metric perturbations at the boundary would also, in analogy with the discussion below for the photon, potentially lead to a dynamical graviton on the boundary.
Formulating the boundary condition for plasmons would in general involve the bulk induction tensor, which makes writing down an explicit expression in terms of $\delta\!A$ somewhat intricate due to the model-dependence.
However, with the advantage of the holographic approach being that it can describe complicated strongly correlated systems by relatively simple gravitational models, we will focus on cases which are commonly used in the literature, where the bulk kinetic term for the $U(1)$ gauge field is usually such that
\begin{eqnarray}
\delta\!\calJ_x	&\propto & 	\delta\!A'_x \, \bigr |_{\partial M}
\, ,
\end{eqnarray}
where prime denotes the normal derivative at the boundary.
In this case, it is a direct consequence of the above discussion that the plasmon boundary condition is of the form,
\begin{eqnarray}
\Bigr(\omega^2 \delta\!A_x+ p(\omega,k) \, \lambda \,\delta\!A_x' \, \Bigr)\,\Bigr|_{\partial M}	&=&	0
~.
\label{eq:RobinPlasmon}
\end{eqnarray}
Here, $p$ is a function that will depend on details of the bulk kinetic term of the gauge field.
For clarity, we have also explicitly mentioned how the scale $\lambda$ from the holographic identifications~\eqref{eq:holodict} enters the boundary condition. This also makes manifest how plasmons are related to a double trace-deformation of the boundary theory, since adding the double-trace potential $W({\cal O}) \sim \lambda {\cal O}^2$ to the boundary field theory leads to a linear relation between the source ($\delta\!A_x$) and the expectation value ($\delta\!A_x'$) of the form (\ref{eq:RobinPlasmon}), whereby the new coupling $\lambda$ has been introduced -- see \cite{Zaanen:2015oix} for further details. Note how the condition (\ref{eq:RobinPlasmon}) is explicitly related to the introduction of a new scale, $\lambda$, while the Dirichlet condition for QNMs would be entirely unaffected.

In most commonly considered models the function $p$ is likely to be bounded, if not even entirely independent of $\omega$ and $k$.
This in general implies that for very small $\omega$, the second term dominates and one has roughly a Neumann boundary condition, but for large $\omega$ the first term dominates and one instead has a Dirichlet boundary condition.
This interpolation is physically reasonable since in the limit of large $\omega$ the rapid oscillations means that the dynamical polarization can be neglected, and hence Dirichlet boundary conditions are appropriate. In the opposite limit of small $\omega$, the photon mediated dynamical polarization is key and it has been argued~\cite{Witten:2003ya} that imposing Neumann boundary conditions leads to a dynamical photon in the boundary QFT.



\section{Holographic Plasmons in the Reissner--Nordstr\"om Metal}
\label{sec:RNmetal}
We now apply the concept outlined in section~\ref{sec:holoplasmons} to the RN metal, i.e.~the field theory dual of an electrically charged Reissner--Nordstr\"om black hole spacetime in $(3+1)$-dimensional anti--de Sitter space (AdS), which is a solution that extremizes the Einstein--Maxwell action,
\bea
S   \;=\;   \int_M * (R-2\Lambda) - \int_M F \wedge *F  \, .
\label{eq:action}
\eea
Note that, as the dual field theory lives on the conformal asymptotic boundary of the AdS bulk spacetime, the RN metal we consider is effectively $(2+1)$-dimensional.
The RN solution can be parametrized as,
\begin{equation}
    L^{-2}\d s^2=-f(z)\d t^2+\left(\frac{z_0}{z}\right)^2\d x^2+\left(\frac{z_0}{z}\right)^2\d y^2+g(z)\frac{\d z^2}{z_0^2}\,,\label{eq:backgroundmetric}
\end{equation}
with
\begin{align}
    f(z)&=\left(\frac{z_0}{z}\right)^2-M \left(\frac{z}{z_0}\right)+\frac{1}{2} Q^2 \left(\frac{z}{z_0}\right)^2,\qquad
    g(z)=\left(\frac{z_0}{z}\right)^4 \frac{1}{f(z)}\,.
\end{align}

Besides the AdS length scale $L$ this spacetime is determined by two parameters, the horizon radius $z_0$ and the charge of the black hole $Q$. The latter is bounded by $0\leq Q^2\leq 6$, the limits correspond to the Schwarzschild and the extremal solution, respectively.
As mentioned in the discussion following~\eqref{eq:holodict}, we have to keep in mind that the boundary field strength $\calF = d\calA$ is, in principle, only fixed up to a choice of scale in terms of $F|_{\partial M}$. With the boundary current $\calJ$ being, by construction, the canonical conjugate to $\calA$ in the grand canonical ensemble with potential given by the Euclidean on-shell action, $\Omega = T S^{Eucl,on-shell}$, and $F=W$ in the bulk, we thus have,
\bea
\calA   \;=\;	\frac{1}{\sqrt{\lambda}}\,  A \Bigr|_{\partial M}	\, , \quad
\calJ 	\;=\;	\sqrt{\lambda}\,A' \Bigr|_{\partial M}	\, ,
\label{eq:holoidentification}
\eea
For our two-parameter family of solutions we can then define temperature, chemical potential and charge density in the dual field theory~\cite{Ammon:2015wua,Zaanen:2015oix,Hartnoll:2016apf},
\begin{eqnarray}
T			\;=\;	\frac{6-Q^2}{8 \pi L z_0}	\, , \quad
\mu			\;=\;	\frac{Q}{\sqrt{\lambda}\, L z_0}	\, , \quad
n	\;=\;	\frac{\sqrt{\lambda}\, Q}{L^2 z_0^2}	\, .
\end{eqnarray}
To analyze the dielectric function, we use linear response, i.e.~we perturb the system around the above mentioned background, considering longitudinal excitations and working in radial gauge.
This results in a system of six coupled differential equations, involving four metric and two gauge field perturbations.
This system is similar to the one obtained when studying QNMs -- see e.g.~\cite{Edalati:2010pn}, which is also how we chose conventions, up to reparametrizing the charge $Q \rightarrow{\sqrt{2} Q}$ and RG flow parameter $r_0/r \rightarrow{z/z_0}$.
The crucial difference to QNM calculations is the choice of boundary conditions at asymptotic infinity which define the holographic dictionary.
For plasmons, this condition has to be the one discussed in section~\ref{sec:holoplasmons}, i.e.~that all perturbations except $\delta\!A_x$ vanish at the boundary.
 In order to obtain a well-posed boundary value problem we also need to supplement boundary conditions at the event horizon, which will be the usual physical requirement that no modes will be emitted by the black hole. This entails two modes satisfying infalling boundary conditions, as well as four pure gauge modes, analogous to calculating QNMs in a non-gauge-invariant parametrization -- see e.g.~\cite{Amado:2009ts}.

 
Finding the plasmon mode requires to find the dispersion relation defined by the plasmon condition~$\varepsilon(\omega,k)=0$. 
This is equivalent to finding the values of $\omega$ and $k$ where~\eqref{eq:RobinPlasmon} is satisfied. With the identification~\eqref{eq:holoidentification} we have $p \equiv 1$ and plasmons in this model are thus characterized by 
\begin{eqnarray}
\left.
\Bigr( \omega^2 \delta\!A_x + \lambda\, \delta\!A'_x \,\Bigr)\,\right|_{\partial M}	&=&	0
~.
\label{eq:BC_Plasmon_RN}
\end{eqnarray}
The points $(\omega,k)$ where non-trivial solutions exist define the plasmon dispersion relation $\omega(k)$. 
Our numerical investigations have revealed that the qualitative behavior of the dispersion curves is hardly affected by the choice of $\lambda$, when keeping $\mu/T$ fixed, so we will only present data with $\lambda=1$. Note however, that the gap for plasmon dispersion, and other quantitative features do depend on the value of $\lambda$. In particular we found that the gap scales linearly with $\lambda$ for $\lambda \ll1$ and levels off to a constant with $\lambda\gg1$.
Further investigation of the quantitative implications of the choice of $\lambda$ will be left for a future publication.
 
\begin{figure}[ht]
    \centering
    \begin{minipage}[t]{0.45\textwidth}
        \centering
        \includegraphics[width=0.9\textwidth]{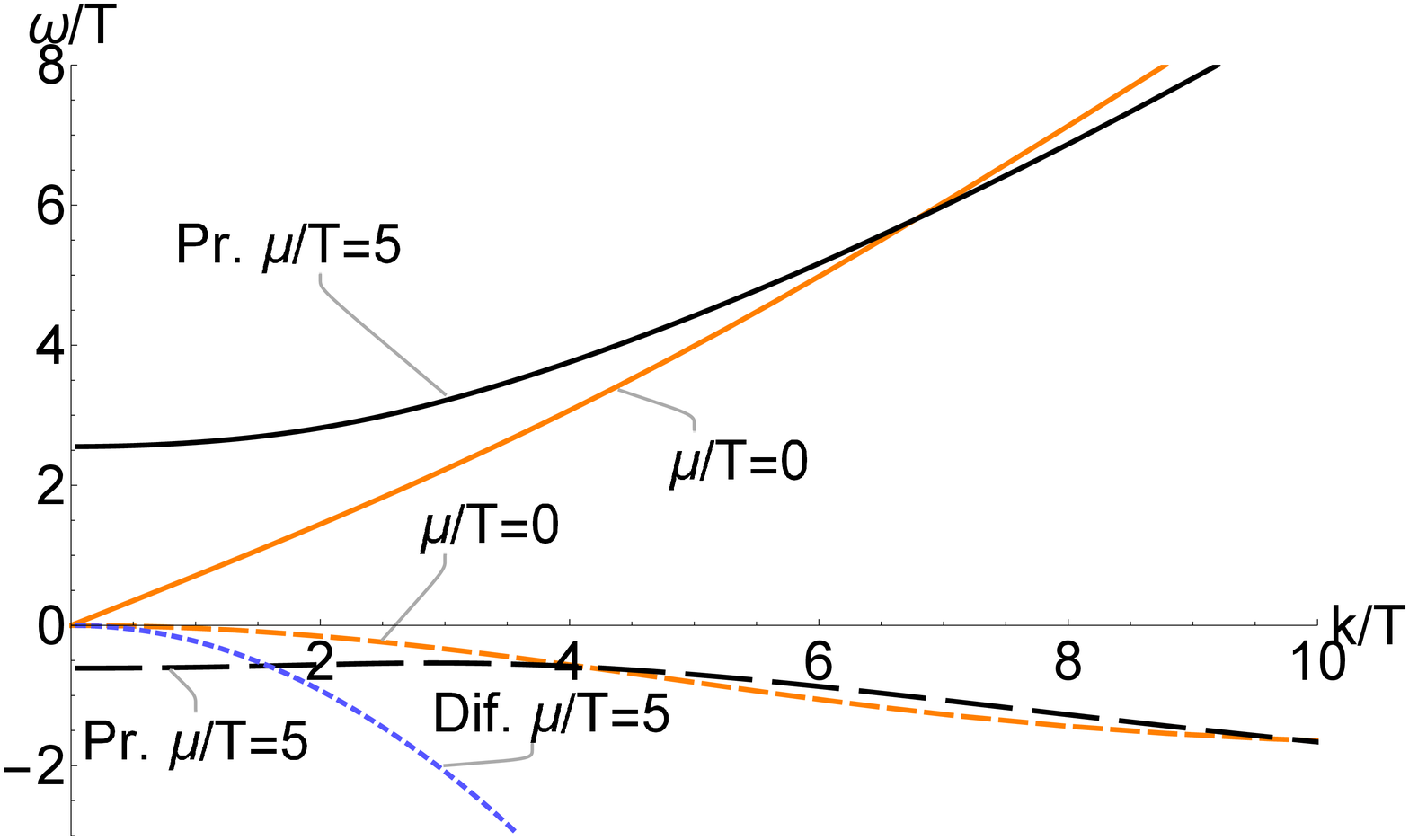} 
        \caption{The lowest ${\varepsilon=0}$ modes for the RN background at $\mu/T=0$ and $\mu/T=5$. Imaginary parts are dashed. Note the sound mode whose imaginary part vanishes as $k/T$ approaches zero and the gapped propagating plasmon mode (Pr.) with initially larger imaginary part than the purely diffusive mode (Dif.).\label{fig:dispersion}}
    \end{minipage}\hfill
    \begin{minipage}[t]{0.45\textwidth}
        \centering
        \includegraphics[width=0.9\textwidth]{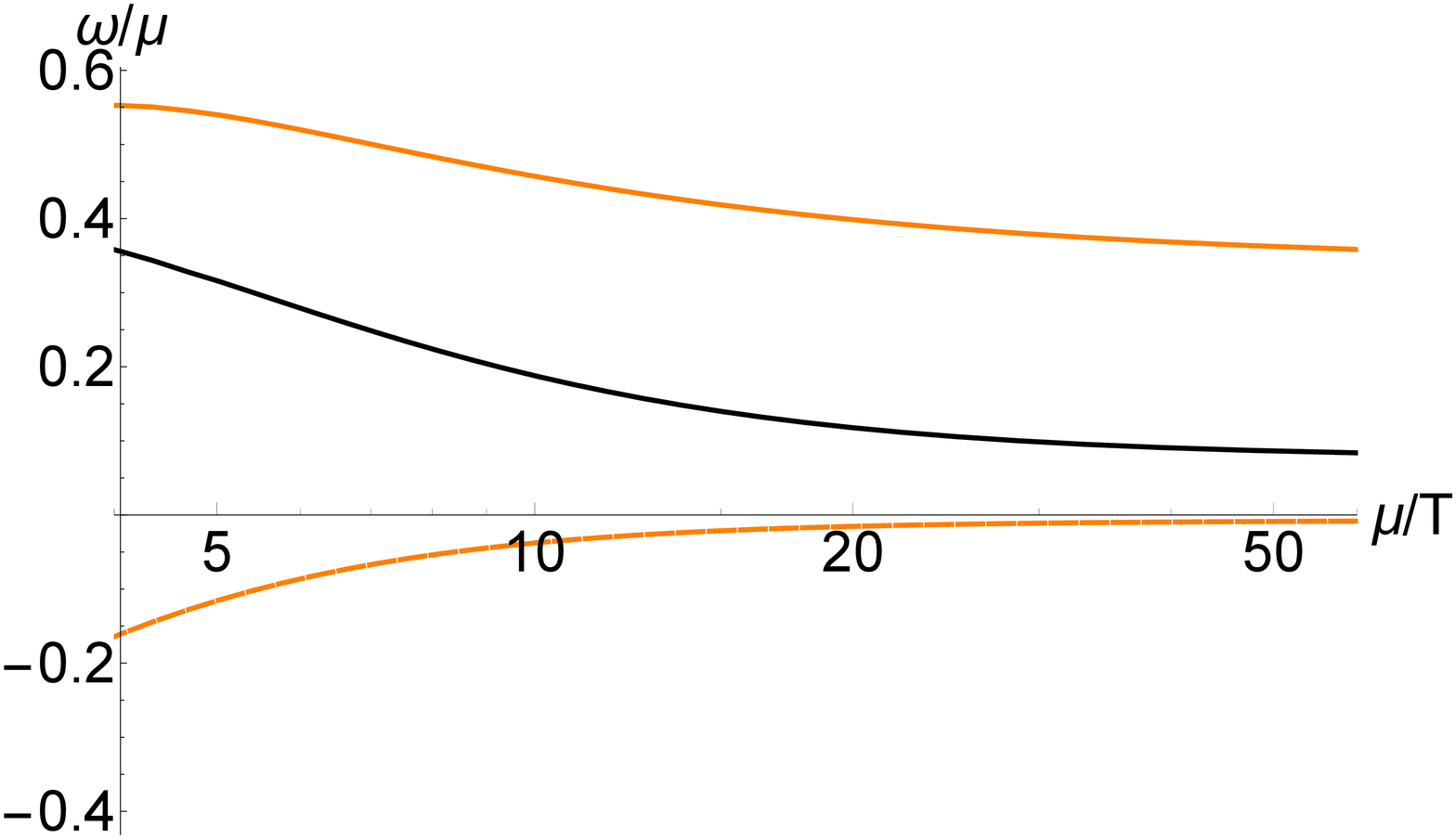} 
        \caption{The crossover between the diffusive and gapped mode at different $\mu/T$. The black line is $k/\mu$, and the dashed orange line is $\mathrm{Im}[\omega]/\mu$, at crossover where the two modes are equally long-lived. The solid orange line shows $\mathrm{Re}[\omega]/\mu$ of the gapped plasmon mode at the crossover.\label{fig:crossover}}
    \end{minipage}\hfill
\end{figure}

The results are presented in figures~\ref{fig:dispersion} and~\ref{fig:crossover}.
There are a number of important features to note here. 
Firstly, consider figure~\ref{fig:dispersion}. For uncharged systems, we see the standard sound mode.
When turning on a charge, this mode becomes gapped and can be identified with the plasmon. 
The gap at $k=0$ has no simple scaling with $\lambda$. For small $k$, we see that the most stable ${\varepsilon=0}$ mode, is a purely diffusive one. The imaginary part of this mode grows quickly, so for slightly larger $k$ the gapped plasmon mode is the most stable one. 
This is consistent with the expectation that the strongly coupled plasmon mode, at non-zero temperature, does not dominate in the hydrodynamic regime close to $k/\mu=0$.

Furthermore, while relatively stable for small $k$, the imaginary part of the gapped plasmon mode starts to grow for larger $k$. This result seems to be in qualitative agreement with experimental results~\cite{2017arXiv170801929M} where the plasmon mode is only seen for small values of $k$. 
It should be stressed, however, that the holographic system is of codimension zero, i.e.~charge and electric field live in the same number of dimensions, as opposed to e.g.~surface excitations, where the electric fields permeate one additional dimension, which corresponds to codimension one.
As a matter of fact, it is the codimension which is the important quantity to be considered, and codimension zero systems in two or three dimensions have similar dispersion relations for plasmons -- this is a direct consequence of the Fourier transform of the Coulomb potential in the appropriate number of dimensions.
Thus, even though we use a $(2+1)$-dimensional boundary, the plasmons are expected to disperse as physical three-dimensional plasmons (sometimes called \emph{bulk} plasmons, not to be confused with the bulk spacetime in holography). Despite the effective number of dimensions in which the Coulomb potential acts being different in the two systems the decreasing lifetime of the plasmon for larger values of $k$ is expected to be a generic feature of plasmons at strong coupling.

In figure~\ref{fig:crossover} we give the value of $k/\mu$ above which the plasmon mode outlives the diffusion mode, as a function of $\mu/T$.
This crossover is identified by equal values of~$\mathrm{Im}[\omega]/\mu$.
As expected, at lower temperatures and fixed chemical potential, the plasmon mode is more long-lived for a larger range of $k/\mu$, and it becomes increasingly stable at crossover. The real part of $\omega/\mu$ at crossover is included for completeness. The result in figure~\ref{fig:crossover} can be viewed as a holographic prediction, to be addressed in future experiments.

Furthermore, for small $\omega$ and $k$ one can compare the holographic results to those following from hydrodynamics. Combining equation (\ref{eq:diel_00}) and (\ref{eq:kubo}), and using the hydrodynamical form of the longitudinal correlator from \cite{Kovtun:2012rj}, for finite density and momentum, one finds that the real part of the plasmon dispersion in figure \ref{fig:dispersion} agrees to lowest non-trivial order in $k/T$. However, the imaginary part of the plasmon dispersion is qualitatively different as it goes to zero as $k/T \rightarrow 0$ from hydrodynamics, while it instead approaches a constant, non-zero, value from holography.

Note however that the plasmon mode is \emph{not} a hydrodynamic mode since it is gapped, and the hydrodynamic data we have used, in the form of series expansions around $\omega=k=0$, might be applied outside its radius of convergence in the comparison above.

Finally, the data from our numerical results also allows for an important consistency check, which is to verify the sum rules for the dielectric function, c.f.~\cite{mahan2000many}. Note that the sum rules probe more information regarding the dielectric function than the dispersion relations displayed in the figures above, which we get from just the poles of~$\epsilon(\omega,k)^{-1}$.
Indeed we find (up to numerical precision) 
\begin{equation}
    \lim_{k\to0}\int_0^\infty\frac{\mathrm{d}\omega}{\omega}\, \mathrm{Im}\,[\epsilon(\omega,k)^{-1}]=-\pi/2\,.
\end{equation}

\section{Conclusions and Outlook}
\label{sec:conclusions}
In this paper we perform the first holographic computation of the dispersion relation for plasmons. The key steps are to first note that the standard definition of the dielectric function~\cite{nozieres1966theory} holds even at strong coupling, and in the absence of quasiparticles. The plasmon condition $\varepsilon(\omega,k) = 0$ in the boundary field theory can then be translated into boundary conditions for bulk fluctuations. 

We use this method to compute the plasmon dispersion for the standard holographic toy model of an RN metal and obtain a gapped plasmon dispersion as expected for codimension zero systems, i.e.~where the extension of the system and the electromagnetic field are the same -- a brief description of plasmons in different number of dimensions can be found in~\cite{mexicanpaper}.
The application of this formalism to more physically realistic holographic models is a natural next step\footnote{U.~Gran, M.~Torns\"o and T.~Zingg, In preparation.}.

An interesting open question is how to model systems with a non-trivial plasmon dispersion in a top-down approach. Note that tuning down the chemical potential on the boundary to zero, the gravitational and electromagnetic differential equations decouple, and one can identify an origin for each of the modes. In particular, the gapped plasmon mode has its origin in the gravitational sector, as the sound mode. This implies that for systems that do not have a very high chemical potential, the gravitational fluctuations are essential to model plasmons.
In particular, the standard probe limit approach will not suffice, since the fluctuations of the induced metric on the probe brane are not enough and one would e.g.~also need a non-zero $\delta\! g_{tx}$.


Having obtained the expected dispersion relation for codimension zero plasmons, being qualitatively the same as for three-dimensional plasmons, it would be interesting to extend this formalism to codimension one systems, i.e.~surface plasmons, and in particular to plasmons in a sheet of graphene. For graphene, one can straightforwardly compute the plasmon dispersion from the two-dimensional conductivity for graphene computed from hydrodynamics~\cite{Muller:2008qx}. One then obtains the expected $\omega \propto \sqrt{k}$ dispersion, to leading order in $k$, for graphene and other two-dimensional materials (or, more generally, codimension one materials). 


To summarize, as recent advances concerning spectroscopy of strange metals~\cite{2017arXiv170801929M} will provide a wealth of new data in the near future, a fundamental understanding of plasmon properties in holographic models of strange metals opens up for new, significant progress in the field.


\acknowledgments

We would like to thank Blaise Goutéraux, Matthias Ihl, Andreas Isacsson, Koenraad Schalm, Henk Stoof and Tobias Wenger for valuable discussions. This work is supported by the Swedish Research Council.

\bigskip
\bibliography{holoplasmons}

\providecommand{\href}[2]{#2}\begingroup\raggedright\begin{thebibliography}{10}

\bibitem{Maier:2007:10.1007/0-387-37825-1}
S.~Maier, \emph{Plasmonics: Fundamentals and applications}.
\newblock Springer US, 2007,
  \href{https://doi.org/10.1007/0-387-37825-1}{10.1007/0-387-37825-1}.

\bibitem{2011NanoL..11.3370K}
F.~H.~L. {Koppens}, D.~E. {Chang} and F.~J. {Garc{\'{\i}}a de Abajo},
  \emph{{Graphene Plasmonics: A Platform for Strong Light-Matter
  Interactions}}, \href{https://doi.org/10.1021/nl201771h}{\emph{Nano Letters}
  {\bfseries 11} (Aug., 2011) 3370--3377},
  [\href{https://arxiv.org/abs/1104.2068}{{\ttfamily 1104.2068}}].

\bibitem{SPSO2003}
W.~L.~Barnes, A.~Dereux and T.~W.~Ebbesen, \emph{Surface plasmon subwavelength
  optics}, {\emph{Nature} {\bfseries 424} (08, 2003) 824--830}.

\bibitem{ZIA200620}
R.~Zia, J.~A. Schuller, A.~Chandran and M.~L. Brongersma, \emph{Plasmonics: the
  next chip-scale technology},
  \href{https://doi.org/https://doi.org/10.1016/S1369-7021(06)71572-3}{\emph{Materials
  Today} {\bfseries 9} (2006) 20 -- 27}.

\bibitem{2017arXiv170801929M}
M.~{Mitrano}, A.~A. {Husain}, S.~{Vig}, A.~{Kogar}, M.~S. {Rak}, S.~I. {Rubeck}
  et~al., \emph{{Singular density fluctuations in the strange metal phase of a
  copper-oxide superconductor}},
  \href{https://arxiv.org/abs/1708.01929}{{\ttfamily 1708.01929}}.

\bibitem{vandeMarel:2003wn}
D.~van~de Marel, H.~J.~A. Molegraaf, J.~Zaanen, Z.~Nussinov, F.~Carbone,
  A.~Damascelli et~al., \emph{{Quantum critical behaviour in a high-Tc
  superconductor}}, \href{https://doi.org/10.1038/nature01978}{\emph{Nature}
  {\bfseries 425} (2003) 271}.

\bibitem{Cooper603}
R.~A. Cooper, Y.~Wang, B.~Vignolle, O.~J. Lipscombe, S.~M. Hayden, Y.~Tanabe
  et~al., \emph{Anomalous criticality in the electrical resistivity of
  {$La_{2\text{\textendash}x}Sr_xCuO_4$}},
  \href{https://doi.org/10.1126/science.1165015}{\emph{Science} {\bfseries 323}
  (2009) 603--607}.

\bibitem{Ammon:2015wua}
M.~Ammon and J.~Erdmenger, \emph{{Gauge/gravity duality}}.
\newblock Cambridge University Press, 2015.

\bibitem{Zaanen:2015oix}
J.~Zaanen, Y.-W. Sun, Y.~Liu and K.~Schalm, \emph{{Holographic Duality in
  Condensed Matter Physics}}.
\newblock Cambridge Univ. Press, 2015.

\bibitem{Hartnoll:2016apf}
S.~A. Hartnoll, A.~Lucas and S.~Sachdev, \emph{{Holographic quantum matter}},
  \href{https://arxiv.org/abs/1612.07324}{{\ttfamily 1612.07324}}.

\bibitem{Maldacena:1997re}
J.~M. Maldacena, \emph{{The Large N limit of superconformal field theories and
  supergravity}}, \href{https://doi.org/10.1023/A:1026654312961}{\emph{Int. J.
  Theor. Phys.} {\bfseries 38} (1999) 1113--1133},
  [\href{https://arxiv.org/abs/hep-th/9711200}{{\ttfamily hep-th/9711200}}].

\bibitem{Gubser:1998bc}
S.~S. Gubser, I.~R. Klebanov and A.~M. Polyakov, \emph{{Gauge theory
  correlators from noncritical string theory}},
  \href{https://doi.org/10.1016/S0370-2693(98)00377-3}{\emph{Phys. Lett.}
  {\bfseries B428} (1998) 105--114},
  [\href{https://arxiv.org/abs/hep-th/9802109}{{\ttfamily hep-th/9802109}}].

\bibitem{Witten:1998qj}
E.~Witten, \emph{{Anti-de Sitter space and holography}}, {\emph{Adv. Theor.
  Math. Phys.} {\bfseries 2} (1998) 253--291},
  [\href{https://arxiv.org/abs/hep-th/9802150}{{\ttfamily hep-th/9802150}}].

\bibitem{Fischetti:2012rd}
D.~Marolf, W.~Kelly and S.~Fischetti, \emph{{Conserved Charges in
  Asymptotically (Locally) AdS Spacetimes}},  in \emph{Springer Handbook of
  Spacetime} (A.~Ashtekar and V.~Petkov, eds.), pp.~381--407.
\newblock Springer, 2014.
\newblock \href{https://arxiv.org/abs/1211.6347}{{\ttfamily 1211.6347}}.
\newblock \href{https://doi.org/10.1007/978-3-642-41992-8_19}{DOI}.

\bibitem{Witten:2001ua}
E.~Witten, \emph{{Multitrace operators, boundary conditions, and AdS / CFT
  correspondence}},  \href{https://arxiv.org/abs/hep-th/0112258}{{\ttfamily
  hep-th/0112258}}.

\bibitem{Mueck:2002gm}
W.~Mueck, \emph{{An Improved correspondence formula for AdS / CFT with
  multitrace operators}},
  \href{https://doi.org/10.1016/S0370-2693(02)01487-9}{\emph{Phys. Lett.}
  {\bfseries B531} (2002) 301--304},
  [\href{https://arxiv.org/abs/hep-th/0201100}{{\ttfamily hep-th/0201100}}].

\bibitem{Wenger2017}
T.~Wenger, \emph{Graphene plasmons in nanostructured environments}.
\newblock PhD thesis. Department of Microtechnology and Nanoscience, Applied
  Quantum Physics, Chalmers University of Technology, 2017.

\bibitem{nozieres1966theory}
D.~Pines and P.~Nozières, \emph{{The Theory of Quantum Liquids}}.
\newblock W.A. Benjamin Inc., 1966.

\bibitem{Amariti:2010jw}
A.~Amariti, D.~Forcella, A.~Mariotti and G.~Policastro, \emph{{Holographic
  Optics and Negative Refractive Index}},
  \href{https://doi.org/10.1007/JHEP04(2011)036}{\emph{JHEP} {\bfseries 04}
  (2011) 036}, [\href{https://arxiv.org/abs/1006.5714}{{\ttfamily 1006.5714}}].

\bibitem{Forcella:2014dwa}
D.~Forcella, A.~Mezzalira and D.~Musso, \emph{{Electromagnetic response of
  strongly coupled plasmas}},
  \href{https://doi.org/10.1007/JHEP11(2014)153}{\emph{JHEP} {\bfseries 11}
  (2014) 153}, [\href{https://arxiv.org/abs/1404.4048}{{\ttfamily 1404.4048}}].

\bibitem{Liu:2015sea}
L.~Liu and H.~Liu, \emph{{Wake potential in a strong coupling plasma from the
  AdS/CFT correspondence}},
  \href{https://doi.org/10.1103/PhysRevD.93.085011}{\emph{Phys. Rev.}
  {\bfseries D93} (2016) 085011},
  [\href{https://arxiv.org/abs/1502.01841}{{\ttfamily 1502.01841}}].

\bibitem{Witten:2003ya}
E.~Witten, \emph{{SL(2,Z) action on three-dimensional conformal field theories
  with Abelian symmetry}},  in \emph{From fields to strings} (M.~Shifman,
  A.~Vainshtein and J.~Wheater, eds.), vol.~2, pp.~1173--1200.
\newblock World Scientific, 2003.
\newblock \href{https://arxiv.org/abs/hep-th/0307041}{{\ttfamily
  hep-th/0307041}}.

\bibitem{Edalati:2010pn}
M.~Edalati, J.~I. Jottar and R.~G. Leigh, \emph{{Holography and the sound of
  criticality}}, \href{https://doi.org/10.1007/JHEP10(2010)058}{\emph{JHEP}
  {\bfseries 10} (2010) 058},
  [\href{https://arxiv.org/abs/1005.4075}{{\ttfamily 1005.4075}}].

\bibitem{Amado:2009ts}
I.~Amado, M.~Kaminski and K.~Landsteiner, \emph{{Hydrodynamics of Holographic
  Superconductors}},
  \href{https://doi.org/10.1088/1126-6708/2009/05/021}{\emph{JHEP} {\bfseries
  05} (2009) 021}, [\href{https://arxiv.org/abs/0903.2209}{{\ttfamily
  0903.2209}}].

\bibitem{Kovtun:2012rj}
P.~Kovtun, \emph{{Lectures on hydrodynamic fluctuations in relativistic
  theories}}, \href{https://doi.org/10.1088/1751-8113/45/47/473001}{\emph{J.
  Phys.} {\bfseries A45} (2012) 473001},
  [\href{https://arxiv.org/abs/1205.5040}{{\ttfamily 1205.5040}}].

\bibitem{mahan2000many}
G.~Mahan, \emph{Many-Particle Physics}.
\newblock Physics of Solids and Liquids. Springer US, 2000.

\bibitem{mexicanpaper}
B.~M. Santoyo and M.~del Castillo-Mussot, \emph{Plasmons in three, two and one
  dimension}, {\emph{Rev. Mex. Física} {\bfseries 4} (1993) 640--652}.

\bibitem{Muller:2008qx}
M.~M{\"u}ller and S.~Sachdev, \emph{{Collective cyclotron motion of the
  relativistic plasma in graphene}},
  \href{https://doi.org/10.1103/PhysRevB.78.115419}{\emph{Phys. Rev.}
  {\bfseries B78} (2008) 115419},
  [\href{https://arxiv.org/abs/0801.2970}{{\ttfamily 0801.2970}}].

\end{thebibliography}\endgroup
\bibliographystyle{JHEP}

\end{document}